\documentclass[preprint]{aastex}
\usepackage{emulateapj5}
\usepackage{apjfonts}
\usepackage{natbib}

\input psfig.tex

\def\aa{{A\&A}}
\def\aas{{ A\&AS}}

\def\al{$\alpha$}
\def\bet{$\beta$}

\def\annrev{{ARA\&A}}
\def\apj{{ApJ}}
\def\apjs{{ApJS}}

\def\cc{cm$^{-3}$}

\def\cc{cm$^{-3}$}

\def\etal{{et al. }}

\def\lamb{$\lambda$}
\def\lax{{$\mathrel{\hbox{\rlap{\hbox{\lower4pt\hbox{$\sim$}}}\hbox{$<$}}}$}}
\def\gax{{$\mathrel{\hbox{\rlap{\hbox{\lower4pt\hbox{$\sim$}}}\hbox{$>$}}}$}}
\def\simlt{\lower.5ex\hbox{$\; \buildrel < \over \sim \;$}}
\def\simgt{\lower.5ex\hbox{$\; \buildrel > \over \sim \;$}}

\def\micron{{$\mu$m}}
\def\mnras{{MNRAS}}

\def\percm2{cm$^{-2}$}

\def\solum{$L_\odot$}
\def\solmass{$M_\odot$}

\def\hii{\ion{H}{2}}
\def\oii{[\ion{O}{2}]}

\def\neii{[\ion{Ne}{2}]}

\def\ne{Ne}
\def\neiii{[\ion{Ne}{3}]}

\slugcomment{To appear in {\it The Astrophysical Journal}.}
\lefthead{Ho \& Keto}
\righthead{Neon Lines and the Star Formation Rate}

\begin{document}

\title{The Mid-infrared Fine-structure Lines of Neon as an Indicator of Star Formation Rate in Galaxies}

\author{Luis C. Ho}

\affil{The Observatories of the Carnegie Institution of Washington, 813 Santa 
Barbara St., Pasadena, CA 91101}

\and

\author{Eric Keto}

\affil{Harvard-Smithsonian Center for Astrophysics, 60 Garden St.,
Cambridge, MA 02138}

\begin{abstract}
The fine-structure lines of singly (\neii\ 12.8 \micron) and doubly (\neiii\ 
15.6 \micron) ionized neon are among the most prominent features in the 
mid-infrared spectra of star-forming regions, and have the potential to be 
a powerful new indicator of the star formation rate in galaxies.  Using a 
sample of star-forming galaxies with measurements of the fine-structure lines 
available from the literature, we show that the sum of the \neii\ and \neiii\ 
luminosities obeys a tight, linear correlation with the total infrared 
luminosity, over 5 orders of magnitude in luminosity.  We discuss the 
formation of the lines and their relation with the Lyman continuum 
luminosity.  A simple calibration between star formation rate and the 
\neii\ +\neiii\ luminosity is presented.
\end{abstract}


\section{Introduction}

The star formation rate (SFR) is one of the most fundamental parameters in 
galaxy evolution, but not trivial to measure.  Practical methods of 
estimating the SFR ultimately rely on indirect indicators of the hydrogen
Lyman continuum luminosity.  From this, one estimates the number of massive, 
ionizing stars, and for an assumed initial mass function, the total stellar 
mass.  The most commonly used SFR estimators make use of the optical hydrogen 
recombination lines and \oii\ \lamb3727 (see, e.g., Kennicutt 1998 for a 
review).  Depending on the application, a variety of estimators based on 
continuum emission have been devised and intercalibrated, including X-rays, 
mid-ultraviolet, far-infrared (IR), and radio synchrotron (e.g., Bell 2003; 
Gilfanov et al. 2004).  

The advent of the {\it Infrared Space Observatory (ISO)}\, and especially the 
{\it Spitzer Space Telescope}\ has brought renewed interest in mid-IR 
diagnostics of star formation.  Two mid-IR SFR estimators have been discussed: 
the strength of polycyclic aromatic hydrocarbon (PAH) features (Peeters et al. 
2004; Wu et al. 2005) and the mid-IR continuum (Roussel et al.  2001; Wu et 
al.  2005).  These estimators, however, are not without complications.  
The PAH features are known to be exceptionally weak in low-metallicity galaxies 
(e.g., Wu et al. 2006, and references therein), their strength being governed 
by complicated, but poorly understood formation and destruction processes 
(O'Halloran et al. 2006; Wu et al. 2006).  PAH emission arises from a 
variety of interstellar environments, not only star-forming sites, and thus 
its ability to reliably trace starburst regions remains a matter of debate 
(Haas et al. 2002; Peeters et al. 2004).  The mid-IR continuum strength, on 
the other hand, shows considerable variation in its relation to the total IR 
luminosity for star-forming galaxies (Dale et al.  2005; but see Takeuchi et 
al. 2005).  Moreover, without additional diagnostics it is often difficult to 
tell whether the primary source of dust heating responsible for the mid-IR 
emission derives from star formation or active galactic nuclei.

This paper proposes a new SFR indicator based on the strength of the 
fine-structure transitions of \neii\ \lamb 12.814 \micron\ ($^2P_{1/2} 
\rightarrow ^2P_{3/2}$) and \neiii\ \lamb 15.554 \micron\ ($^3P_{1} 
\rightarrow ^3P_{2}$).  \neii\ is an excellent tracer of ionizing
stars because (1) its ionization potential (21.56 eV) is higher than that of 
hydrogen; (2) it is a dominant ionization species in \hii\ regions, being 
one of the principal coolants (Burbidge et al. 1963; Gould 1963); (3) its 
relatively high critical density 
($n_{\rm c} = 4.3\times10^5$ \cc; Petrosian 1970) ensures
that the line flux is insensitive to density, since most \hii\ regions have 
electron densities substantially below $n_{\rm c}$; (4) it is abundant 
(solar Ne/H = $1.20\times10^{-4}$; Grevesse \& Sauval 1998), with a 
nucleosynthetic history closely following that of oxygen (Garnett 2002);
and (5) its long wavelength significantly reduces its sensitivity to 
dust extinction compared to optical lines (the extinction at mid-IR 
wavelengths is only a few percent of the extinction at optical wavelengths).
Doubly ionized \ne\ shares these properties, with the difference that \neiii\
can be the dominant ionic species in regions of lower density ($n_{\rm e} < 
100$ cm$^{-3}$) and higher excitation (blackbody radiation temperatures 
$T_{\rm r} > 4\times10^4$ K).  For example, low-mass, low-metallicity 
galaxies, which are known to exhibit systematically higher excitation, show a 
greater fraction of ionized neon emission in the \neiii\ 15.6 \micron\  line 
rather than in the \neii\ 12.8 \micron\ line (O'Halloran et al. 2006; Wu 
et al. 2006).

We demonstrate, using a sample of star-forming galaxies with mid-IR spectra, 
that the sum of  \neii\ + \neiii\ luminosity tightly correlates with a 
well-known measure of star formation, the total IR luminosity.
We discuss the ionization of Ne and the connection between the 
\neii\ and \neiii\ emission and Lyman continuum luminosity, from which a 
simple relationship between \neii\ + \neiii\ luminosity and SFR follows.

\section{Data}

To explore the utility of the mid-IR fine-structure lines of \ne\ as a SFR 
indicator, we have assembled from the literature data for star-forming 
galaxies with measured \neii\  and \neiii\ fluxes.  We have strived to ensure 
that the data were obtained with a 
sufficiently large aperture so that most of the emission is 
%
%
included.  Since 
the \neii\ line sits adjacent to the prominent 12.7 \micron\ PAH feature, it 
is also important that the spectra have sufficient spectral resolution so that 
the line can be unambiguously resolved. Lastly, while the resulting compilation
is necessarily heterogeneous, we have tried to minimize systematic effects by 
concentrating only on a few relatively large, uniformly analyzed samples from 
the literature that were observed with the same instrumental setup.  The final 
sample of 57 sources, summarized in Figure~1, contains 34 galaxies observed 
with the Short Wavelength Spectrometer (SWS) on {\it ISO}\ (Genzel et al. 
1998; Thornley et al. 2000; Verma et al. 2003) and 23 galaxies observed with 
the Infrared Spectrograph (IRS) on {\it Spitzer}\ (O'Halloran et al. 2006; Wu 
et al. 2006).  Although some of the sources are well-known composite systems 
that contain both an active galactic nucleus and a starburst (e.g., Arp~220 
and NGC~6240), it has been previously established that the bulk of the 
bolometric luminosity in these sources is dominated by the starburst component.
Our diverse sample spans a wide range of galaxy types, from low-mass, 
low-metallicity dwarf galaxies to massive merger remnants.  Among the 57 
objects in the sample, 16 are early-type spirals (Sa--Sbc; $\langle M_B \rangle 
= -19.4$ mag), 7 are late-type spirals (Sc--Sm; $\langle M_B \rangle = -18.8$ 
mag), 14 are irregulars and blue compact dwarfs (Im, I0, BCD; $\langle M_B 
\rangle = -17.1$ mag), and 20 are well-known merger or strongly interacting 
systems with peculiar morphologies ($\langle M_B \rangle = -20.4$ mag).

Figure~1 illustrates that the \neii\  emission strongly correlates with the IR 
emission, over 5 orders of magnitude in luminosity.  An ordinary least-squares 
regression of $L_{\rm IR}$ on $L_{\rm [Ne~{\sc II}]}$ yields

\begin{equation}
\log L_{\rm [Ne~{\sc II}]} \ = \ (1.01 \pm 0.054)\log L_{\rm IR} \ - \ 
(3.44 \pm 0.56),
\end{equation}

\vskip 0.2cm
\noindent
with a scatter of 0.51 dex.  The correlation appears to steepen toward low 
luminosities ($L_{\rm IR}$ \lax\ $10^{9.5}$ \solum), in that less \neii\ 
emission is produced for a given amount of IR luminosity. 
%
%
The lower-luminosity
galaxies have a higher ratio of \neiii\ to \neii\ emission, suggesting that in 
these galaxies a significant fraction of the Ne is doubly ionized, most likely
because in a low-metallicity environment the ionizing radiation field 
hardens as a result of reduced line blanketing and blocking in stellar 
atmospheres (e.g., Thornley et al. 2000; Madden et al. 2002).  We examine this 
effect by plotting the sum of \neii\ and \neiii\ (solid points in 
Fig.~1), for those objects for which both lines are measured.  Indeed, the 
lower-luminosity points now exhibit significantly less scatter and follow more 
closely a smooth continuation of the trend defined by the higher-luminosity 
systems.  An ordinary least-squares regression of $L_{\rm IR}$ on 
$L_{\rm [Ne~{\sc II}]+ [Ne~{\sc III}]}$ yields

\begin{equation}
\log L_{\rm [Ne~{\sc II}]+ [Ne~{\sc III}]} \ = \ 
(0.98 \pm 0.069)\log L_{\rm IR} \ - \ (2.78 \pm 0.70)
\end{equation}

\vskip 0.2cm
\noindent
and a scatter of 0.49 dex.  

The IR luminosity is a robust indicator of the SFR in star-forming galaxies 
because a significant fraction of the bolometric luminosity of young stars 
is absorbed and reradiated by dust grains.  We expect the IR and Ne 
luminosities to be related because, as discussed in the next section, the 
lines of ionized \ne\  directly trace the Lyman continuum, which is itself 
produced primarily by young stars.  The tight correlation between the 
luminosity of ionized \ne\ and the IR luminosity indicates that the luminosity 
of the \ne\ fine-structure lines can be used as an effective substitute to 
estimate the SFR.  The theoretical basis for this expectation is examined in 
the next section.

\section{The Relation between the Neon Lines and SFR}

\subsection{The Ionization of \ne}

The relative abundances of the ionic species are set by a balance of the rates 
of photoionization $P_{\rm i}$ and recombination $R_{\rm i}$ for each species 
$N_{\rm i}$:

\begin{equation}
N_{\rm i} \ P_{\rm i} = N_{\rm i+1} \ R_{\rm i} \ n_{\rm e}.
\end{equation}

\vskip 0.2cm
\noindent
The photoionization rate is the integral over a frequency-dependent absorption
cross section $\alpha_{\rm i}(\nu)$ and the ionizing radiation $J(\nu)$ above 
the ionization threshold $\nu_{\rm t}$:

\begin{equation}
P_{\rm i} = \int^\infty_{\nu_{\rm t}} {{4\pi}\over{h\nu}} \ J(\nu) \ 
\alpha_{\rm i}(\nu) \ d\nu.
\end{equation}

\vskip 0.2cm
\noindent
In the case of Ne, Figure~2 shows that the absorption cross section of Ne$^{+}$
peaks at a frequency corresponding to a Wien 
%
%
temperature of $1.5\times 10^5$ K,
which is higher than the characteristic radiation temperature of early-type 
stars ($T_{\rm r} \approx 4\times 10^4$ K).  Thus, in a region ionized by 
typical massive stars the dominant ionic species would be  Ne$^{+}$.  However, 
as shown in Figure~3, \neiii\ will begin to replace \neii\ in ionized gas that 
is at lower density or subject to intense or high-temperature radiation. The 
ionization equilibrium shown in Figure~3 is calculated in the absence of a 
specific model by expressing the mean radiation as diluted blackbody 
radiation,

\begin{equation}
4\pi J_\nu = \pi \Gamma B_\nu,
\end{equation}

\vskip 0.2cm
\noindent
where $\Gamma$ is analogous to the $r^{-2}$ geometrical dilution in a spherical 
model. In Equation 3, the dilution factor may then be combined as a ratio, 
$\Gamma/n_{\rm e}$, with the density, which is also unspecified in the absence 
of a specific model.  Figure~3 illustrates how the use of the combination of 
\neii\ + \neiii\ improves the robustness of \ne\ as an indicator of the SFR,
by extending its applicability to galaxies with higher $\Gamma/n_e$, for 
example, to lower-luminosity, lower-metallicity galaxies that have higher 
excitation (e.g., Thornley et al. 2000; Madden et al. 2006).

\subsection{The Line Intensity}

The general expression for the intensity of a spectral line is, 

\begin{equation}
I \ ({\rm{ergs}\ \  \rm{s}^{-1}\  \rm{cm}^{-2}\ \rm{sr}^{-1}}\ ) = 
{{h\nu \ A_{21}}\over{4\pi}} \int n_2 \ d\ell ,
\end{equation}

\vskip 0.2cm
\noindent
where $A_{21}$ is the Einstein coefficient for spontaneous emission, $n_2$ 
is the number density of ions in the upper state, and the integral is over 
pathlength $d \ell$. The population of the upper state is in general found from
the equilibrium set by the rate equations for collisional and radiative 
transitions between all levels of a species.  The radiative rates are 
considerably faster than the collisional rates at electron densities below
the critical density for collisional de-excitation, $\sim 10^5$ cm$^{-3}$, 
for both the \neii\ and \neiii\ fine-structure transitions. Therefore, for 
lower densities, it is possible to approximate the rate of emission, 
$A_{21}n_2$, by the upward collisional rate alone. In the case of the \neii\ 
doublet, the line intensity is then approximately

\begin{equation}
I_{\rm [Ne~{\sc II}]} = {{h\nu}\over{4\pi}} \gamma \ f_{+} \ C_{12} \ 
EM  ,
\end{equation}

\vskip 0.2cm
\noindent
where $\gamma$ is the interstellar abundance of neon relative to hydrogen 
($9.9\times10^{-5}$; Simpson et al. 1998), $ f_{+}$ is the fraction of 
\ne\ that is singly ionized (Fig.~3), and $C_{12} \ n_{\rm e}$ is the upward 
collisional rate (s$^{-1}$) between the $^2P_{3/2}$ and the $^2P_{1/2}$ levels, 
and $EM = \int n_{\rm e}^2 \ d\ell$ is the emission measure. 

In the case of the \neiii\ triplet, one needs to consider the upward 
transitions to both higher states because a collisional transition to the 
uppermost $^3P_0$ level will result in a radiative transition to the 
intermediate $^3P_1$ level, followed immediately by a transition at 
\lamb 15.6 \micron\ to the ground state $^3P_2$. The rate of the electric
quadrupole transition  $^3P_0 \rightarrow ^3P_2$  is negligible compared to the 
rates of the two magnetic dipole transitions $(\Delta J = 1)$.  Thus, the 
intensity of \neiii\ is approximately 

\begin{equation}
I_{\rm [Ne~{\sc III}]} = {{h\nu}\over{4\pi}} \gamma \ f_{+2} \ 
(C_{12} + C_{13}) \ EM  ,
\end{equation}

\vskip 0.2cm
\noindent
where the subscripts 1, 2, and 3 on the collision rates refer to the levels 
$^3P_2$, $^3P_1$, and $^3P_0$, respectively, in order of increasing energy.

The collision rates are related to the collision strengths $\Omega_{21}$,
assuming a thermal velocity for the electrons, 

\begin{equation} 
C_{12} = 8.63\times 10^{-8} \ {{\Omega_{21} }\over { \omega_1 }} \ 
\left({{10^4~\rm{K} }\over{T_{\rm e}}} \right)^{1/2} ,
\end{equation}

%
%
%
\vskip 0.2cm
\noindent
where $\omega_1$ is the statistical weight of the lower level. For the \neii\ 
\lamb12.8 \micron\ line, $\Omega_{21}/\omega_1 = 0.0785$ (Griffin \etal 2001).
For the \neiii\ \lamb15.6 \micron\ line, $\Omega_{21}/\omega_1 = 0.1548$
and $\Omega_{31}/\omega_1 = 0.0042$ (Butler \& Zeippen 1994).

Integrating over the source area $A=\int d\Omega$ and assuming $T_{\rm e} 
\approx 10^4$ K, the sum of the luminosities of the two lines can be expressed 
as

\begin{eqnarray}
L_{\rm [Ne~{\sc II}]+ [Ne~{\sc III}]} \ ({\rm{ergs}\ \  \rm{s}^{-1}}\ ) = 
3.16\times10^{42} 
\nonumber
\end{eqnarray}
\begin{eqnarray}
\times \left(f_{+} + 1.67 f_{+2}\right) \left({{D_{\rm L}}\over{{\rm Mpc}}}\right)^2
\left({{A}\over{{\rm sr}}}\right)
\left({{EM}\over{{\rm cm}^{-6}\ {\rm pc}}}\right).
\end{eqnarray}
\vskip 0.3cm

The Ne luminosity can be easily related to the Lyman continuum luminosity
by recognizing that, in equilibrium, the ionizing rate of hydrogen is equal to 
the recombination rate:

\begin{equation}
N_{\rm ion} \ ({\rm photons} \ \ {\rm s}^{-1}\ ) = 
n_{\rm e} n_{\rm p} \alpha_{\rm B} V = \alpha_{\rm B} \ EM \ A,
\end{equation}

\vskip 0.2cm
\noindent
where $\alpha_{\rm B} = 2.6\times10^{-13}$ ${\rm cm}^3~{\rm s}^{-1}$ is the 
Case B recombination rate and $V$ is the volume of the emitting region. 
Combining Equations 10 and 11,

\begin{eqnarray}
L_{\rm [Ne~{\sc II}]+ [Ne~{\sc III}]} \ ({\rm{ergs}\ \  \rm{s}^{-1}}\ )  = 
4.15\times10^{-13} 
\nonumber
\end{eqnarray}
\begin{eqnarray}
\times \left(f_{+} + 1.67 f_{+2}\right) \ f_{\rm ion} \ N_{\rm ion} \ ({\rm photons}\ \  {\rm s}^{-1}).
\end{eqnarray}

\vskip 0.2cm
\noindent
Here, $f_{\rm ion}$ is the number fraction of ionizing photons that are 
actually absorbed by the gas (as opposed to being absorbed by dust or freely 
escaped from the nebula).  Hirashita et al. (2003) estimate that 
$f_{\rm ion}=0.6\pm0.2$ in starburst galaxies.

To check the validity of Equation 12, we present in Figure~4 a compilation of 
\neii, \neiii, and Br\al\ measurements for a large collection of \hii\ regions 
observed with the {\it ISO}\ SWS.  The data for the \hii\ regions in the 
Galaxy and the Small and Large Magellanic Clouds come from Giveon et al. 
(2002), while the M33 sources come from Willner \& Nelson-Patel 
(2002)\footnote{Giveon et al. (2002) also give data for some M33 \hii\ regions,
but the data set of Willner \& Nelson-Patel (2002) is more extensive.}.  
Three points deserve notice.  First, for the Galactic sources both 
\neii\ and the sum of \neii\ + \neiii\ trace Br\al\ linearly.  Second, for
the extragalactic sources, all of which have subsolar abundances 
(especially the Small Magellanic Cloud), \neii\ systematically deviates 
from a linear correlation with Br\al\ whereas the sum of \neii\ + \neiii\ 
remains linear.  Finally, the relation given in Equation 12, with 
$f_{+} = 0.75$ (which is a reasonable estimate from the peak 
of the curves shown in Fig.~3) and $f_{\rm ion} = 0.6$, extremely well 
describes the \neii\ data for the Galactic sources.  The same relation, 
with a small adjustment in zeropoint to account for the ionization fraction, 
very well matches the sum of \neii\ + \neiii\ for {\it all}\ the sources; the 
scatter around the linear relation is only $\sim 0.26$ dex, which is expected 
simply from the metallicity variation among the four galaxies and within each 
galaxy (e.g., Willner \& Nelson-Patel 2002).


Given the relation between the \neii\ + \neiii\ luminosity and the ionizing 
photon rate, it is straightforward to translate 
$L_{\rm [Ne~{\sc II}]+ [Ne~{\sc III}]}$ into a SFR, using one of 
the many existing calibrations between SFR and $N_{\rm ion}$.  Adopting the 
calibration of Kennicutt (1998), which assumes solar abundances and a 
Salpeter (1955) initial mass function with a lower mass limit of 
0.1 \solmass\ and an upper mass limit of 100 \solmass, and choosing a fiducial 
value of $f_{\rm ion} = 0.6$,

\begin{equation}
{\rm SFR} \ (M_{\odot}\ {\rm yr^{-1}}) = 4.34\times10^{-41}  \ 
\left[ { { L_{\rm [Ne~{\sc II}]+ [Ne~{\sc III}]} \ ({\rm ergs} \ \ {\rm s}^{-1}) }\over{ f_{+} + 1.67 f_{+2} } } \right] .
\end{equation}

\vskip 0.6cm

\section{Discussion and Summary}

We have investigated the utility of the mid-IR fine-structure lines of Ne as a 
SFR indicator.  Being dominant coolants in \hii\ regions, the 
\neii\ and \neiii\ lines are among the most prominent and easy-to-measure 
features in the mid-IR spectra of virtually all extragalactic sources, 
especially with the advent of {\it Spitzer}.  Unlike commonly used SFR 
indicators such as the synchrotron radio continuum or even the IR luminosity, 
which relate to massive stars in a complex manner, the formation of the \neii\ 
and \neiii\ lines is particularly simple.  Their intensity scales with the 
emission measure of an ionized nebula, and hence directly with the Lyman 
continuum luminosity.  This property makes them an extremely attractive tool 
to estimate the SFR in extragalactic systems.  The obvious advantage of 
\neii\ and \neiii\ over optical lines such as H\al\ and \oii\ is their greatly 
reduced sensitivity to dust extinction.

Using a sample of star-forming galaxies with available mid-IR spectra from the 
literature, we show that the \neii\ luminosity correlates linearly with the
total IR luminosity, over 5 orders of magnitude in luminosity and ranging 
from low-metallicity dwarf galaxies to high-metallicity, ultraluminous 
IR galaxies.   The observed scatter in this correlation is only $\sim 0.5$ dex, 
which is remarkably tight, considering several likely sources of variation.
First, the neon abundances span at least a factor of 5 (e.g., Verma et al. 
2003).  Second, aperture effects must contribute to at least part of the 
scatter.  Both the SWS aperture and the effective aperture of the IRS (either 
in mapping or staring mode) are significantly smaller than the {\it IRAS}\ 
beam.  Although one usually expects the IR emission to be centrally peaked in 
starburst systems, given the diversity of our sample it is quite likely that 
the IR emission arises from a larger area than is sampled by the \neii\ 
measurements.  The data set from Thornley et al. (2000) attempted to correct 
the IR emission for aperture effects, but in general this was not available 
for the other data sets.  To estimate the likely impact of aperture effects, 
we sorted the sample by optical angular diameter, but find
no obvious dependence of angular size scatter with scatter around the 
correlation.  Third, the fractional ionization of Ne is variable. 
However, inclusion of both the \neii\ and \neiii\ luminosities improves the 
empirical correlation in that the scatter and turnover of the correlation at 
the low-luminosity end decrease with the combined luminosities.  Lastly, while 
the extinction at mid-IR wavelengths is much less than in the optical, in some 
cases it can still be non-negligible (e.g., Genzel et al. 1998).  

A quantitative explanation of the $L_{\rm [Ne~{\sc II}]+ [Ne~{\sc III}]}$ vs. 
$L_{\rm IR}$ correlation is nontrivial, since the IR emission arises not only 
from dust grain heating by massive stars but also by nonionizing stars, both 
young and old (e.g., Hirashita et al. 2003).  However, to a first 
approximation this correlation indicates that, to the extent that the IR 
luminosity traces the SFR (e.g., Kennicutt 1998; Bell 2003; Hirashita et al.  
2003), so, too, does the \neii\ + \neiii\ luminosity, as expected from the
basic theoretical considerations discussed in \S~3.  

To our knowledge, this is the first systematic analysis of the \neii\ and 
\neiii\ lines for a large sample of star-forming galaxies with the expressed
intent of using them to derive SFRs.  In their {\it ISO}\ survey of 
ultraluminous IR galaxies, Genzel et al. (1998) had explicitly recognized that 
the \neii\ luminosity can be used as a surrogate for the Lyman continuum 
luminosity.  From a subset of 12 starburst galaxies, they empirically 
determined that the average ratio of Lyman continuum luminosity to \neii\ 
luminosity is 64.  This value is smaller than that deduced from our 
calculations in \S 3: according to the starburst synthesis models presented in 
Hirashita et al. (2003), the average Lyman continuum photon has an energy of 
$\sim 19.4$ eV, and if $f_{+} = 0.75$, $f_{+2} = 0$, and $f_{\rm ion} = 0.6$, 
then Equation 12 implies $\langle L_{\rm Lyc}/L_{\rm [Ne~{\sc II}]}\rangle 
\approx 166$.  In a recent {\it Spitzer}\ IRS study of low-redshift quasars and 
starburst-dominated ultraluminous IR galaxies, Schweitzer et al.~(2006) 
presented a correlation between \neii\ luminosity and 60 \micron\ luminosity, 
showing that the two quantities are significantly correlated.  Although 
Schweitzer et al.~argue that the \neii\ emission in their sample largely 
originates from star-forming regions rather than the accretion-powered 
narrow-line region, this is difficult to prove in the absence of a 
detailed analysis of the narrow-line spectra of their sample.  In any event, 
our analysis spans a larger dynamic range in luminosity and galaxy environment,
and it conclusively demonstrates that in star-forming galaxies the combined
\neii\ + \neiii\ luminosity tracks the IR emission.

We propose that the relationship between the \neii\ + \neiii\ luminosity and
the SFR (Eq.~13) be added to the arsenal of SFR indicators presently used in 
the literature.  If possible, the \neii\ and \neiii\ luminosities employed in 
Equation 13 should be corrected for extinction.  We note, however, that our 
calibration itself, which is derived from theoretical considerations, does 
{\it not}\ depend on extinction correction.  Our calibration is based on (1) 
the relation between the luminosity of the neon lines and the ionizing photon 
rate (Eq.~12) derived in this paper and (2) a previously determined and widely 
accepted conversion between the ionizing photon rate and the SFR (Kennicutt 
1998).   While the neon-based SFR calibration explicitly depends on the 
fractional abundance of singly and doubly ionized Ne (as expressed by the 
factors $f_{+}$ and $f_{+2}$ in Eq.~13), which can only be ascertained from 
detailed modeling of individual sources, if most of the Ne is ionized, the 
distribution between the \neii\ and \neiii\ states introduces a variation in 
the calibration of less than a factor of 2.

As an independent check of the reliability of our new method, we have 
compared the SFR obtained from the \neii\ + \neiii\ luminosity with the SFR 
derived more traditionally from the IR luminosity for the sample of 57
galaxies plotted in Figure 1.  Using the IR calibration given in Kennicutt 
(1998), the ratio of the IR-based SFRs to the Ne-based SFRs has an average 
value of $1.9\pm3.7$ and a median value of 0.7.  Within the significant 
scatter, there is no evidence for a large systematic bias in the Ne-based 
method.

\acknowledgements
This work was supported by the Carnegie Institution of Washington and by NASA 
grants from the Space Telescope Science Institute (operated by AURA, Inc., 
under NASA contract NAS5-26555).  We thank Alex Dalgarno for discussions 
on collision rates.  We are grateful to an anonymous referee for a 
thoughful and constructive review of the paper.

\clearpage
\begin{figure}
\plotone{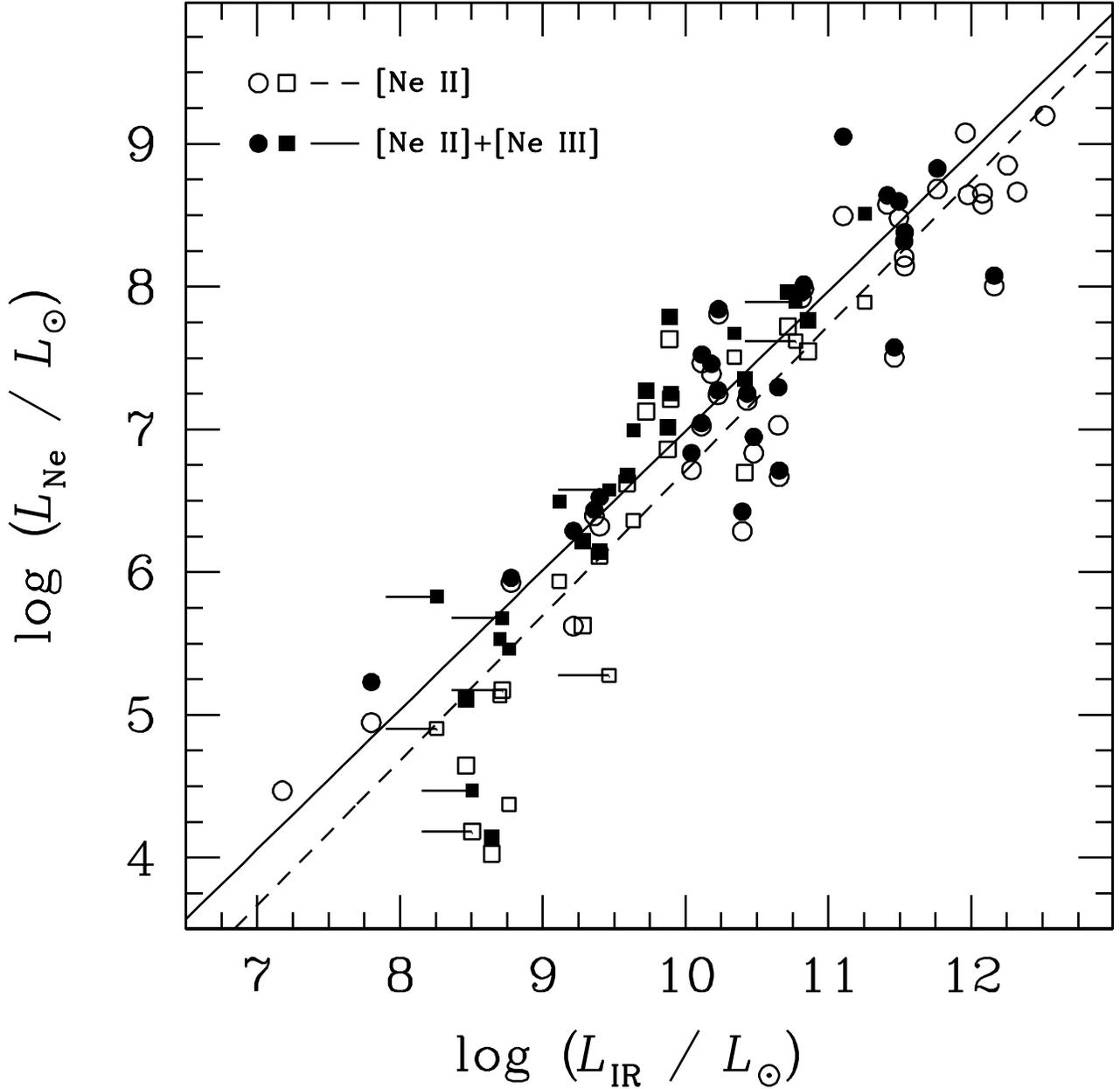}
\caption{
Empirical correlation between ionized neon emission and
the total IR (8--1000 \micron) luminosity, derived using 12, 25, 60, and 100
\micron\ flux densities from the {\it Infrared Astronomical Satellite (IRAS)}
following the prescription of Sanders \& Mirabel (1996).  The open
points denote \neii\ 12.8 \micron\ measurements, whereas the solid points
represent the sum of \neii\ and \neiii\ 15.6 \micron, with the linear
regression fits shown as dashed and solid lines, respectively.  The
measurements were taken from Genzel et al. (1998; excluding the active
galactic nuclei), Thornley et al. (2000), Verma et al.  (2003), O'Halloran et
al. (2006), and Wu et al. (2006).  We have used the distances
adopted by these authors.  Data taken with {\it ISO}\ SWS are plotted
as circles; data taken with {\it Spitzer}\ IRS are plotted as squares; five
sources have upper limits on $L_{\rm IR}$.  If both {\it ISO}\ and
{\it Spitzer}\ data are available, preference is given to the latter.
}
\end{figure}

\clearpage
\begin{figure}
\plotone{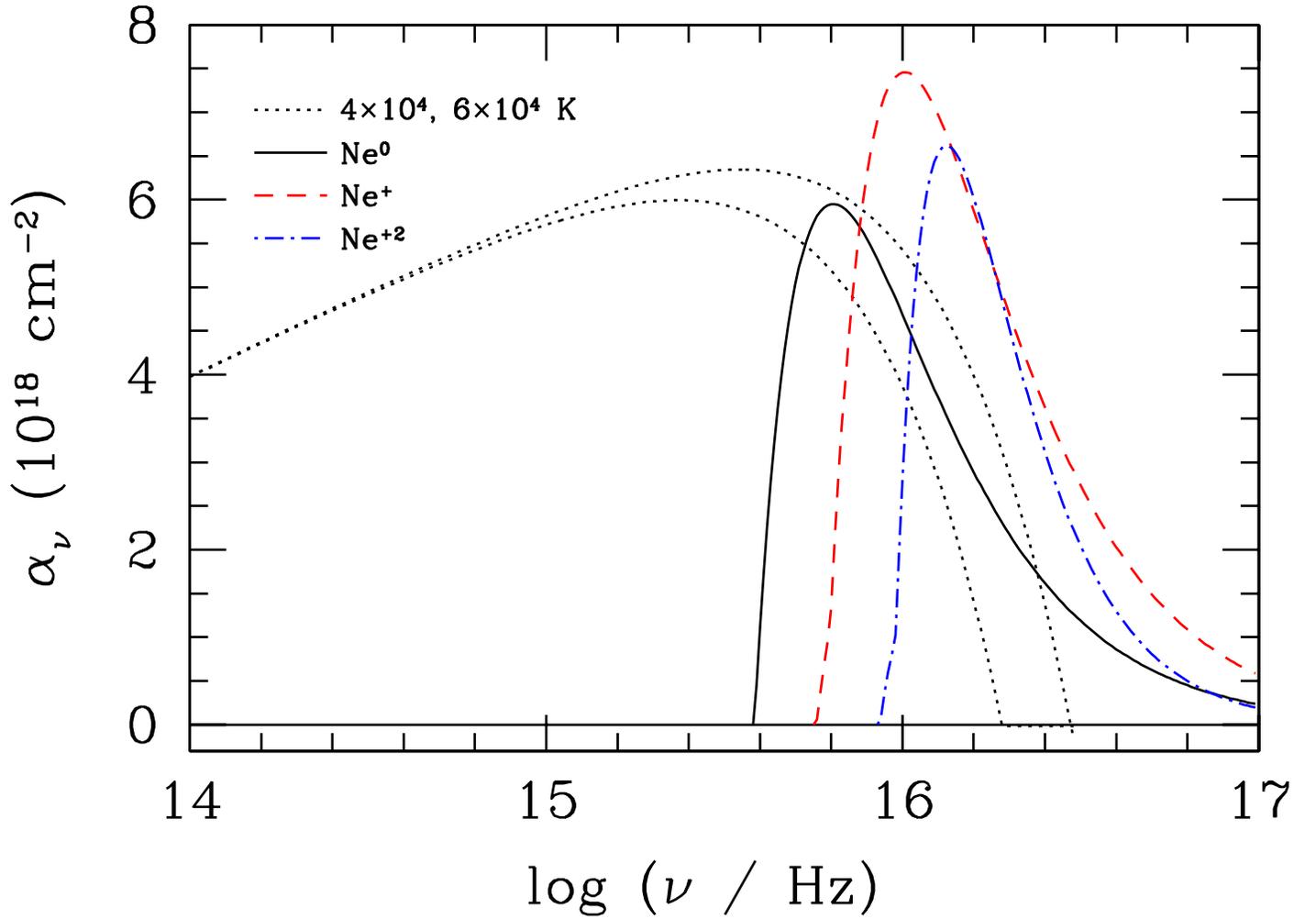}
\caption{
Photoionization cross sections for Ne$^{0}$ ({\it solid, black}),  Ne$^{+}$
({\it dashed, red}), and  Ne$^{+2}$ ({\it dot-dashed, blue}).  The cross
sections are computed using Equation 2.31 and Table 2.7 of Osterbrock (1989).
Shown for comparison are curves for blackbody emission at $4\times 10^4$ and
$6\times 10^4$ K ({\it dotted}), which have been scaled arbitrarily.
}
\end{figure}

\clearpage
\begin{figure}
\plottwo{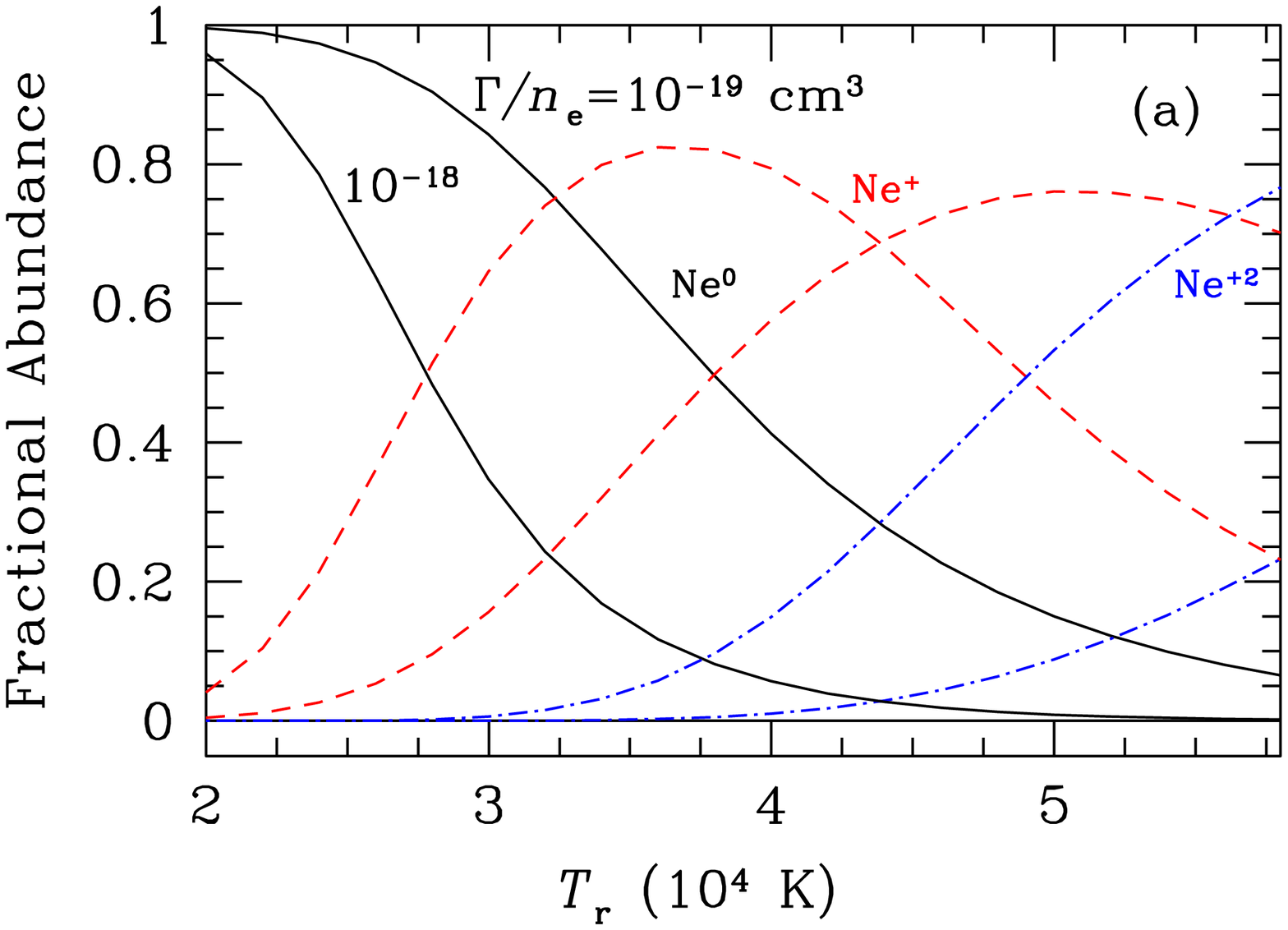}{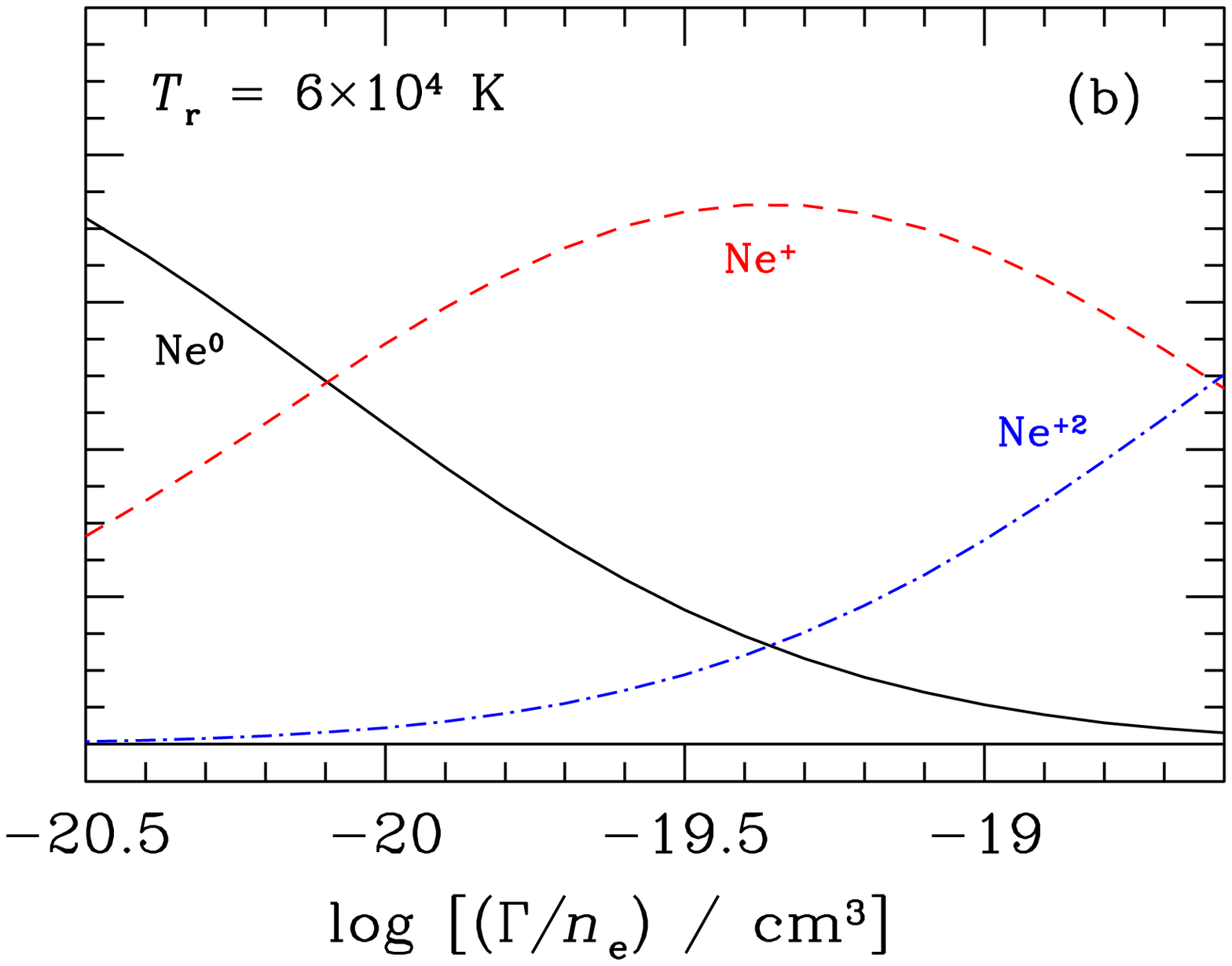}
\caption{
Fractional abundance of Ne$^{0}$ ({\it solid, black}),  Ne$^{+}$ ({\it dashed,
red}), and  Ne$^{+2}$ ({\it dot-dashed, blue}) as a function of ({\it a})
radiation temperature for two fixed values of the ratio of the dilution factor
$\Gamma$ for the blackbody radiation and the density $n_e$ ($\Gamma/n_{\rm e}
= 10^{-19}$, right set of curves; $10^{-18}$ cm$^3$, left set of curves) and
({\it b}) the ratio $\Gamma/n_{\rm e}$ for a fixed radiation temperature of
$T_{\rm r} = 6\times 10^4$ K.  The calculations utilize the photoionization
cross sections from Figure~2, radiative recombination rates from Gould (1978),
and dielectronic recombination rates from Nussbaumer \& Storey (1987). The
recombination rates assume a fiducial temperature of $10^4$ K and that the
population of each species is in the ground state. The photoionization rate
of \neii\ includes transitions to the metastable states of \neiii\ ($^1D$
and $^1S$).  We have ignored charge-exchange transitions involving the
exchange of ionization state between neutral H and the Ne ions, since the
density of neutral H should be negligible in ionized gas.
}
\end{figure}

\clearpage
\begin{figure}
\plotone{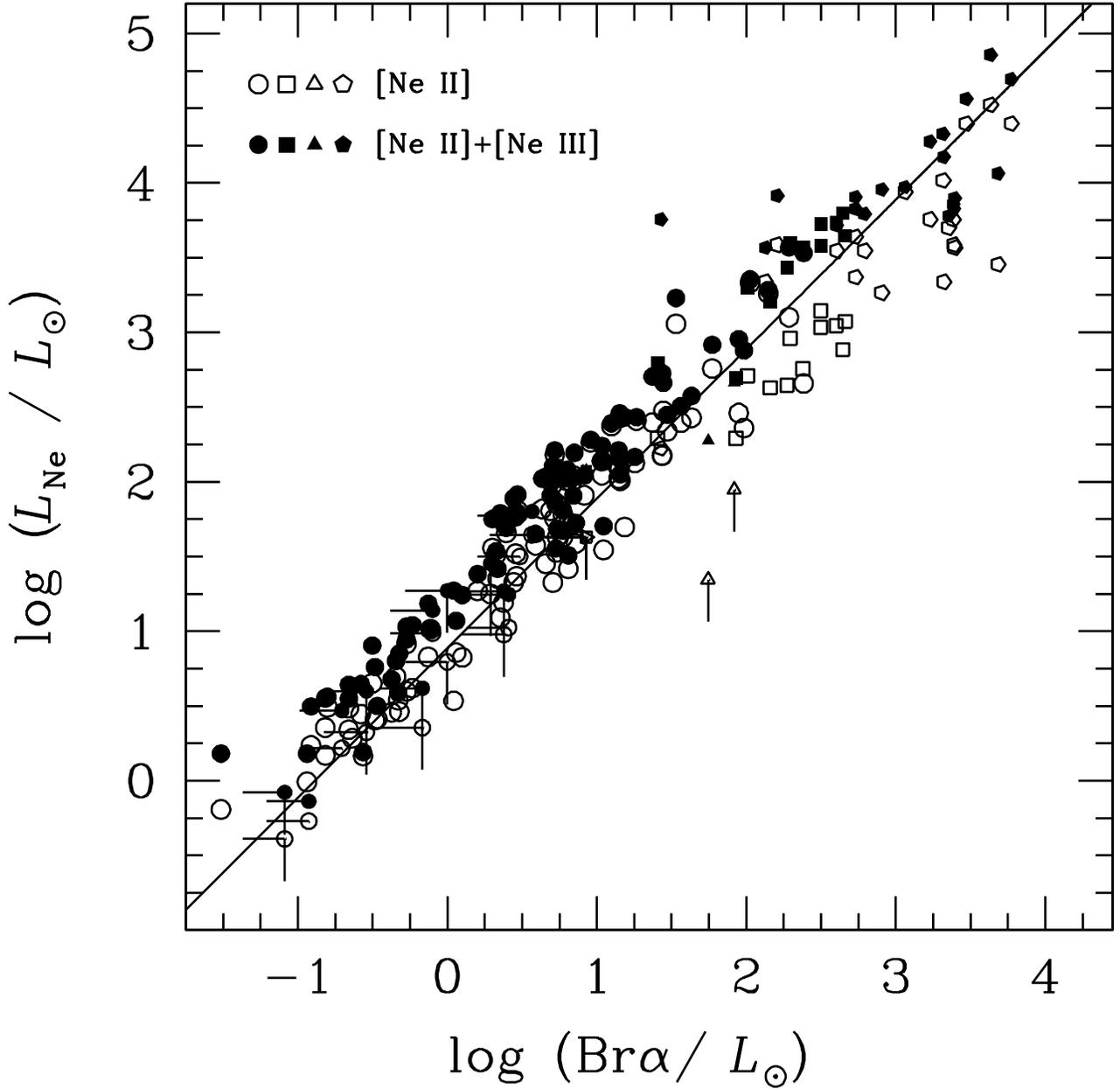}
\caption{
Empirical correlation between ionized neon emission
and the Br\al\ emission for \hii\ regions (Giveon et al. 2002; Willner \&
Nelson-Patel 2002) in the Galaxy ({\it circles}), the Large Magellanic Cloud
({\it squares}), the Small Magellanic Cloud ({\it triangles}), and M33
({\it pentagons}).  The Br\bet\ fluxes for M33 were translated to
Br\al\ assuming a Bracket decrement of 1.6.  The {\it solid line}\
shows the relation given in Equation 12, expressed in terms of Br\al\
using the effective recombination rate for Br\al\ for Case B
recombination ($n_{\rm e} = 10^2$ \cc, $T_{\rm e} = 10^4$ K; Osterbrock 1989)
and assuming $f_{+} = 0.75$ and $f_{\rm ion} = 0.6$.
}
\end{figure}
\end{document}